\documentstyle[a4,epsfig]{article}
\begin{document}
\title{The scalar field-theoretical Coulomb problem}
\author{N.E. Ligterink$^a$\thanks{E-mail: ligterin@ect.it} 
and B.L.G. Bakker$^b$\thanks{E-mail: blgbkkr@nat.vu.nl}\\
{\it $^a$ECT*, Strada delle Tabarelle 286, I-38050 Villazzano (Trento), 
Italy},\\
{\it $^b$ Vrije Universiteit, Department of Physics and Astronomy,}\\
{\it Amsterdam, The Netherlands }
}
\date{\today}
\maketitle
\begin{abstract}
We analyze the fully relativistic, field-theoretical treatment of the scalar 
Coulomb problem. We work in a truncated Hilbert-Fock space containing the
two-constituent states and the 
two-constituent-and-one-massless-exchange-particle
states. Self-energy contributions are dominant for large values of the
coupling constant. Using the covariant formulation of the self-energy does 
not alter these results significantly.  Both the weak-coupling limit and 
the heavy-mass limit lead to the non-relativistic results. 
\end{abstract}
%
\section{Introduction}
In this paper we will solve the simplest possible scalar field-theoretical
Coulomb problem in a Hamiltonian framework \cite{Hei54}. This approach
is much simpler than most other field-theoretical approaches, and
yields better results.  Furthermore, we also include the self-energy
parts, which are usually ignored, and find that they give significant
contributions. Our main motivation for this work lies in improving our
understanding of the field-theoretical bound-state problem before we
turn our attention to stronger binding, and more physical, cases.

There is a wealth of relativistic bound-state equations
\cite{Tod71,BS66,Gro69}.  Most of them are three-dimensional reductions
of the Bethe-Salpeter equation \cite{BS51} in the ladder approximation.
Since there is a degree of arbitrariness involved in this 
reduction \cite{Gro82,Taco95}, the equations can be tailored to satisfy unitarity
\cite{BS66}, gauge invariance \cite{PW98}, or the one-body limit
\cite{Gro82}, which are some of the typical problems with the
underlying Bethe-Salpeter equation \cite{IZ80,Nak69}. However, these
are approximations upon an equation with problems, which is a
questionable practice.

Generally, it is difficult to favour one equation above the other
\cite{STG99}.  Therefore, there has been some attempt to establish
standard results \cite{ST93,Taco95}, which include all orders in
perturbation theory, in the quenched approximation.  To do so one
starts from the path integral for the two-particles four-point function and
integrate out some degrees of freedom analytically, and the resulting
integral is integrated numerically in the Feynman-Schwinger
representation by the Monte Carlo method. The typical problems with this
method are related to the discretisation of space-time, the Wick
rotation, which leads to decaying states of which only the
lowest-energy states can be determined accurately, and the problem to
handle massless particles satisfactorily \cite{Taco95}. However, the
results are promising and show the underbinding of the Bethe-Salpeter
equation in the ladder approximation \cite{NT96,Sav01}.

Another, distinct approach is based on Haag's expansion of asymptotic
states in field theory in free and bound states, which lead to an
hierarchy of coupled equations in the coupling constant expansion
\cite{GRS95}. With some difficulty the scalar field theory, which is
also analyzed here, is solved. The results seem to indicate a failure
to obtain both the one-body limit and the weak-coupling limit in the
Haag-expansion approach.

Most prominently has been the return to Hamiltonian field theory
\cite{Hei54}.  Because the bound-state is a stationary state, and 
manifest covariance is lost, it seems appropriate to solve
the bound-state problem in the Hamiltonian framework. Most calculations
are done in the light-front Hamiltonian framework
\cite{BPP98,BHM99}, which preserves part of the boost symmetry at the
cost of rotational symmetry \cite{Dir49}. We will use the ordinary
instant-form Hamiltonian instead to show that that indeed yields the
same results as the Kepler problem \cite{Flu71}, in the weak-coupling
limit.

The reason to study the case of massless exchange particles in
particular are manifold. Firstly, the massless particles are a typical
and hard problem for field theory, as they generate long-range
interactions, and lead to singularities in the S-matrix associated with
multiple soft-photon emissions \cite{BN37,Wei95}. None of these
problems have to be present in the bound-state calculation.  
Indeed, in the Hamiltonian approach the infrared singularity is absent,
contrary to the nonrelativistic treatment using the Coulomb potential and
the covariant approach. When
constituents are bound, they are off-shell, which means they do not
share enough energy among them to become on-shell.  Therefore they do
certainly not have enough energy to radiate off soft photons, which is
the origin of the infrared complications in the covariant approach
\cite{BN37,JR76}.

Consider the $S$-matrix element of two constituents exchanging a
massless particle. This matrix element has a singularity.  It occurs
because at zero momentum exchange the difference in energy of the
initial and intermediate states vanishes.  If the two constituents
belong to a bound state, the singularity is hidden by the binding
energy.  Apparently, the binding energy has the same effect as the mass
of the exchanged particle: screening at long distances.

Secondly, eventually, one has to admit that all fundamental
interactions that generate bound states are massless, gauge theories.
For the case of positronium, a relativistic approach can be used
\cite{Mur88}, however, rather in the cumbersome Coulomb gauge
\cite{AMF97} where the lowest-order kernel reduces to an instantaneous
interaction, and many other cancellations occur.  For the heavy-light
case, like the hydrogen atom, the starting point has to be the
non-relativistic Coulomb potential \cite{EGS00}, or the equivalent
Dirac equation with such a potential.  This Coulomb potential has
survived many revolutions in physics, mainly because of its
phenomenological success: it predicts atomic spectra to a high accuracy
\cite{BS57,EGS00}. This success stood in the way of fundamental
development, where the classical action-on-a-distance concept is
replaced by the correct notion of particle exchange \cite{BS51,Gro93}.
There is no viable alternative which reproduces the particular
long-range interaction generated by the exchange of massless particles.
At least, there is no alternative that does not introduce additional
problems in the one-body limit. 

In this paper we recover the typical scaling associated with the
weak-coupling Coulomb ground state.
This paper is not just a formal exercise to link theory with
experiment.  Eventually, we like to understand how relativistic effects
come about in deeply bound systems, and give sensible answers to where
the spin of a system is coming from, how the radius of the bound state
is related to the masses and the coupling strength, what the mass of a
constituent is, and which are the effective degrees of freedom.  We
emphasize {\it deeply bound states}, i.e., states with a binding energy
of the same order as the masses of the constituents, because we feel
that for them the problems we want to address are most pressing. For
weakly bound states one can treat relativistic aspects, like
relativistic kinematics and the creation and annihilation of particles
as a correction to the classical approach, where a potential picture is
appropriate. 
Although it would be nice to understand better the origin
and limitations of the potential picture.

The Hamiltonian framework is conceptually quite different from the
standard, covariant field-theoretical framework, although the final
formulae may be similar.  In this framework exchange diagrams exist,
but not as a potential.  Rather, they are to be interpreted as matrix
elements of the fundamental, local interactions, between different Fock
sectors.

In the fundamental description of bound states there is no place for 
the concept of a potential. All interactions are local. What seems to be a 
non-local interaction is due to the smearing of the constituent 
particles, which are not just single particles but a linear 
combination of states with different numbers of bare particles. 

The Hamiltonian description gives us a consistent way to truncate the
infinite complexity arising in field theory, by truncating Fock space
to two- and three-particle states. Many arguments concerning
bound-states are devoted to the truncation to a finite number of
particles, since, in order to generate a bound state which arises only
non-perturbatively, an infinite summation of diagrams is required. So
the question is which finite set one should take to iterate infinitely
many times.  For the Bethe-Salpeter equation \cite{BS51} the results
depend crucially on the answer to that question. However, formulated as
a Hamiltonian eigenvalue problem this infinite summation appears
naturally, given the Fock-state truncation. If this truncation to the
two- and three-particle Fock-state sector is legitimate anywhere, it
should be here, because in the weak-coupling limit a few particles are
expected to contribute and anti-particles (Z-diagrams, which appear in
a covariant approach) are expected to be suppressed, due to their large
virtuality.  We shall analyze the validity of these arguments.

So, for our problem we assume that our bound state consists of two
pieces: one with two oppositely charged scalar particles and one with an
additional, massless scalar particle that can be emitted and absorbed by each
of the charged particles. We use scalar particles to keep the problem as
simple as possible, and free from questions associated with gauge symmetry.
A careful analysis of the three-particle state is
required in order to recover all of the physics. One finds an exchange
contribution and a self-energy contribution; the latter is infinite and
has to be renormalized \cite{Col84}. Of course, we face the problem of
the interpretation of divergent integrals.

The concept of self-energy is still a mystery to many. In solid-state
physics one can accept that a moving electron deforms the crystal,
generating a drag which can be interpreted as an effective mass. In
deep-inelastic scattering one can interpret the self-energy as the
mixing-in of multi-particle states as sufficient energy becomes
available when the particle goes off-shell. In a bound state the
missing energy is the binding energy, which does not belong to a
particular particle, but affects each of the particles separately. One
may look upon this in the following way: each constituent is not a
single particle; partly it is one particle, partly it is two, and so
on. The relative weight of each of these states of a constituent is
determined by two things: the number of states available and the
virtuality of these states $\omega_i$. In a classical world only the
one-particle state $\omega_0$ would exists because that state has the
lowest energy $\omega_0<\omega_i$. In the quantum world and within the
Hamiltonian framework, energy is only conserved asymptotically, so
higher energy states, when available, will also be occupied. The
weights of virtual states is determined by their virtualities. If we
now put a ``compound'' constituent in a bound state with binding energy
$E_b$, the relative virtuality
$\omega_i/\omega_0>(\omega_i+E_b)/(\omega_0+E_b)$ of the multi-particle
states in the constituent is less than for the free particle, therefore
the summed energy will lower, and the self-energy will increase the binding.

\section{Theory}

In the present paper we solve the scalar Coulomb problem. We limit ourselves
to this case in order to avoid numerous problems concerning gauge theories.
Here we are rather interested in the conceptual problem
of what a bound state is and what determines its properties in the simplest
possible case. Spin effects and questions concerning gauge invariance can
be dealt with later.

Our model Lagrangian describes the interaction of three scalar fields
$\phi_a$, $\phi_b$, and $\chi$,
\begin{equation}
 {\cal L}  = 
 - \frac{1}{2} \psi_a 
 [\partial^\mu \partial_\mu +m^2_a + e_a \chi ]\psi_a
 - \frac{1}{2} \psi_b
 [\partial^\mu \partial_\mu +m^2_b + e_b \chi ]\psi_b 
 - \frac{1}{2} \chi \partial^\mu \partial_\mu \chi \, ,
\end{equation}
which have the respective masses: $m_a$, $m_b$, and zero. It will allow
for a bound-state solution, if the charges, $e_a$ and $e_b$, with which
the massive fields couple to the massless field $\chi$ are imaginary and
have opposite signs. This choice makes the Lagrangian non-Hermitian. 
However, also with real coupling constants this Lagrangian
would be ``sick,'' since then the spectrum is not bounded from below
\cite{Bay59} .
However, for a bound state we restrict ourselves to low orders in
particle number, or coupling constant, which, we will show, leads to
sensible results, apparently unaffected by the formal problems with the
Lagrangian.  This Lagrangian would lead also to the Wick-Cutkosky model
\cite{WC54} as only the product $e_a e_b$ would appear.  The only
difference is the opposite sign of the self-energy part, which is not
present in the original W-C model. For real couplings of the same sign,
the problem is similar to scalar gravitation where the same charges
attract, which is precisely the reason why the vacuum is unstable, as
more and more charges together have a lower and lower energy. The
opposite imaginary charges would correspond to electrodynamics, which
would, if implemented properly, require vector coupling and gauge
fields. However, to recover the Balmer spectrum this is not necessary,
and we want to keep the discussion clear by avoiding issues concerning
gauge symmetry. 
However, a more important point to make is that, unlike in
the W-C model, the heavy-mass limit, or one-body limit, where one
constituent has a infinite mass yields the correct solution
\cite{Gro93,EGS00}. Although the spectrum of the W-C model yields the
Balmer series in the weak-coupling limit, the equation does not reduce
to an effective one-body equation, where the heavy particle plays the
role of a static source.

The transition from the Lagrangian formalism to the Hamiltonian one is
rather tedious, but can be found in the literature \cite{Gro93,Hei54}. Here we
only like to make some general comments.  In order to give the theory a
particle interpretation the free fields are quantized. Therefore, if
a field $f(k)$ is known on a space-like hypersurface in 
four-dimensional space-time, it is known in the whole space, via the
mass relation:

\begin{equation}
\int d^4 k \theta_\tau(k) \delta(k^2-m^2) f(k) = \int \frac{d^3 {\mathbf{k}} }
{ \phi_{m}({\mathbf{k}})} f(\omega_{m}({\mathbf{k}}),{\mathbf{k}}) \, ,
\end{equation}

where the phase-space factor $\phi_m$ depends on the mass $m$ and the
three-momentum ${\mathbf{k}}$ perpendicular to the time direction $\tau$.
The on-shell energy $\omega_{m}({\mathbf{k}})$ is the solution of the
equation $k^2-m^2=0$. The $\theta_\tau$ function restricts the integral
to the positive energy solutions. Negative energy solutions are
associated with holes or backward-in-time moving particles, which, in
the Hamiltonian formulation, are the charge- and parity-reversed
partners, and should be treated as distinct particles.

Each intermediate particle is associated with a phase-space factor
$\phi^{-1}$, which, for a time-reversal invariant formulation, should
be divided among the creation and annihilation vertices of this particle.
We will work in the ordinary equal-time formulation, so $\tau = x^0$ and
\begin{eqnarray}
\omega_m({\mathbf{k}}) & = & \sqrt{m^2 + {\mathbf{k}}^2} \, , \\
\phi_m({\mathbf{k}}) & = & 2 \omega_m({\mathbf{k}}) \, .
\end{eqnarray}
For the simplest treatment of the field-theoretical problem above we
assume that the bound state consists of only two different Fock states:
\begin{equation}
|\psi \rangle = |{\mathbf{p}}_a {\mathbf{p}}_b \rangle  +
                |{\mathbf{k}}_a {\mathbf{k}}_b {\mathbf{k}}_\gamma \rangle \, .
\end{equation}
No further approximations are required.  The Hamiltonian restricted to
this subspace of the Hilbert space allows only for the possible
absorption or emission of one massless particle:
\begin{equation}
H = \left( \begin{array}{cc} \omega_2 & \lhd \cr
\rhd & \omega_3 \end{array} \right) \,  ,
\end{equation}
where the diagonal entries are the on-shell energies of the respective
Fock states, and the  off-diagonal entries are the absorption and
emission of the $\chi$ field:
\begin{eqnarray}
 \omega_2 & = & \omega_a({\mathbf{p}}_a) + \omega_b({\mathbf{p}}_b) \, ,  \cr
 \lhd     & = & e_b\big\{^{{\mathbf{p}}_a}_{{\mathbf{k}}_a}\big\} 
\big[^{{\mathbf{p}}_b}_{{\mathbf{k}}_b{\mathbf{k}}_\gamma}\big] +
 e_a\big\{^{{\mathbf{p}}_b}_{{\mathbf{k}}_b}\big\} 
 \big[^{{\mathbf{p}}_a}_{{\mathbf{k}}_a{\mathbf{k}}_\gamma}\big] \, ,\cr
 \rhd     & = &  e_b\big\{^{{\mathbf{k}}_a}_{{\mathbf{p}}_a}\big\} 
 \big[^{{\mathbf{k}}_b {\mathbf{k}}_\gamma}_{{\mathbf{p}}_b}\big]  +
 e_a\big\{^{{\mathbf{k}}_b}_{{\mathbf{p}}_b}\big\} 
\big[^{{\mathbf{k}}_a {\mathbf{k}}_\gamma}_{{\mathbf{p}}_a}\big]\, ,\cr
 \omega_3 & = & \omega_a({\mathbf{k}}_a) + \omega_b({\mathbf{k}}_b) +
 \omega_\gamma({\mathbf{k}}_\gamma) \, .
\end{eqnarray}
Here we introduce $[\cdots]$ as the shorthand notation for 
the normalized Hamiltonian interaction
\begin{eqnarray}
 \big[^{{\mathbf{k}}_a {\mathbf{k}}_b}_{{\mathbf{p}}_c }\big]  & = &
 \int \frac{d^3 {\mathbf{p}}_c 
 \delta^3({\mathbf{k}}_a + {\mathbf{k}}_b - {\mathbf{p}}_c) }
 {\sqrt{\phi_{m_a}({\mathbf{k}}_a) \phi_{m_b}({\mathbf{k}}_b)
 \phi_{m_c}({\mathbf{p}}_c) }} \, , \\
 \big[^{{\mathbf{p}}_a }_{{\mathbf{k}}_c {\mathbf{k}}_d}\big]  & = &
 \int \frac{d^3 {\mathbf{k}}_c  d^3 {\mathbf{k}}_d
 \delta^3({\mathbf{p}}_a - {\mathbf{k}}_c -{\mathbf{k}}_d)}
 {\sqrt{\phi_{m_a}({\mathbf{p}}_a) 
 \phi_{m_c}({\mathbf{k}}_c) \phi_{m_d}({\mathbf{k}}_d)}} \, ,
\end{eqnarray}
and $\{\cdots \}$ generates the proper change of variables for the particles
not affected by the interaction:
\begin{equation}
 \big\{^{{\mathbf{k}}_a}_{{\mathbf{p}}_b}\big\} =
 \int {d^3 {\mathbf{p}}_b} \delta^3({\mathbf{k}}_a-{\mathbf{p}}_b) \, .
\end{equation}
We solve the eigenvalue equation above by introducing the eigenvalue
$E$ as a Lagrange multiplier, and expressing the three-particle state
in terms of the two-particle state and a function of the unknown $E$, from
the eigenvalue equation:
\begin{equation}
|{\mathbf{k}}_a {\mathbf{k}}_b {\mathbf{k}}_\gamma \rangle = \frac{
e_b\big\{^{{\mathbf{k}}_a}_{{\mathbf{p}}_a}\big\} 
\big[^{{\mathbf{k}}_b {\mathbf{k}}_\gamma}_{{\mathbf{p}}_b}\big]  +
e_a\big\{^{{\mathbf{k}}_b}_{{\mathbf{p}}_b}\big\} 
\big[^{{\mathbf{k}}_a {\mathbf{k}}_\gamma}_{{\mathbf{p}}_a}\big] }
{E-\omega_a({\mathbf{k}}_a) - \omega_b({\mathbf{k}}_b) - \omega_\gamma({\mathbf{k}}_\gamma)} 
|{\mathbf{p}}_a {\mathbf{p}}_b \rangle \, .
\end{equation}
This three-particle state shows the interesting features of a
bound-state problem. Although a third particle is present, and so the
number of degrees of freedom has increased, the actual degrees of
freedom are still those of the two-particle wave function, and an
additional parameter in the form of $E$. This wave function has
significant values only in restricted areas of the ${\mathbf{k}}_a\otimes
{\mathbf{k}}_b\otimes {\mathbf{k}}_\gamma$ space; however, large back-to-back
values can occur, unquenched by the two-particle wave function.

The dimensionful coupling constants 
$i e_a/\sqrt{m_am_b}=-i e_b/\sqrt{m_am_b}=e$ are related to the 
fine-structure constant $\alpha = {e^2}/(16 \pi)$, in such a way 
that the bound state is neutral. 
In order to find the bound state we use two sets of variational wave
functions, tuned by the parameter $\mu$:
\begin{eqnarray}
|{\mathbf{p}},{\mathbf{-p}}\rangle_{\rm cov} &  = & \frac{1}{\sqrt{\omega_a \omega_b}
(\mu-m_a-m_b + \omega_a + \omega_b)^2} \, ,\cr 
|{\mathbf{p}},{\mathbf{-p}}\rangle_{\rm nr} &  = & \frac{1}{(2 m \mu +
 {\mathbf{p}}^2)^2} \, , \label{eq.wf}
\end{eqnarray}

where $m$ is the reduced mass.
We will work in the center-of-mass frame of the two constituents,
therefore, $\omega_a =\omega_a ({\mathbf{p}})$ and $\omega_b
=\omega_b(-{\mathbf{p}}) = \omega_b({\mathbf{p}})$. The second wave function
is motivated by the non-relativistic ground-state wave function, while
the first wave function would follow from a relativistic vertex in
Hamiltonian form, after it is differentiated with respect to the scale
parameter $\mu$. In practice the form of wave function does not
matter, although the actual value of the parameter $\mu$ does. The
energies differ only 0.1\% for the different choices of wave
functions.

\begin{figure}
\centerline{\includegraphics[width=10cm]{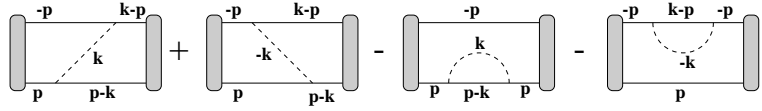}}
 \caption{\label{fig.00} A graphical representation of Eq.~\ref{eq.dia}}
\end{figure}

For each bound-state energy $E$ the coupling constant $\alpha$ is
determined by minimizing it with respect to the wave-function scale
parameter $\mu$:
\begin{equation}
\alpha(\mu) =
\frac{ E \langle {\mathbf{p}},{\mathbf{-p}} | {\mathbf{p}} ,{\mathbf{-p}}\rangle - \langle {\mathbf{p}},{\mathbf{-p}} |\omega_2 | {\mathbf{p}},{\mathbf{-p}} \rangle}
{\alpha^{-1} \langle {\mathbf{p}},{\mathbf{-p}}| \lhd |{\mathbf{k}}_a {\mathbf{k}}_b {\mathbf{k}}_\gamma\rangle}\ \ \ .
\end{equation}
Note that the quantity in the denominator is actually independent of
$\alpha$. It contains two types of contributions: exchange terms and self-energy terms,
\begin{equation}
\langle {\mathbf{p}},{\mathbf{-p}}| \lhd |{\mathbf{k}}_a {\mathbf{k}}_b {\mathbf{k}}_\gamma\rangle   =
  \langle {V_{ab}(E)} \rangle +\langle  {V_{ba}(E)}\rangle 
- \langle \frac{\Sigma_a(E)}{\omega_a} \rangle - \langle  \frac{\Sigma_b(E)}{\omega_b} \rangle
\label{eq.dia}
\end{equation}
where $\langle \cdots \rangle$ indicates the expectation value for the wave functions  $\psi$
of Eq.~\ref{eq.wf}. Equation \ref{eq.dia} is graphically represented in Fig.~\ref{fig.00}.
The exchange part  is finite 
\begin{equation}
\langle {V_{ab}(E)} \rangle = \frac{\alpha m_a m_b}{4 \pi^2}
\int \frac{\psi^\ast({\mathbf{p}})d^3 {\mathbf{p}} }{\sqrt{\omega_a({\mathbf{p}}) \omega_a({\mathbf{p}})}}
\int \frac{d^3 {\mathbf{k}}}{\sqrt{\omega_a({\mathbf{p}}-{\mathbf{k}})
\omega_b({\mathbf{p}}-{\mathbf{k}})}
\omega_\gamma({\mathbf{k}})} \frac{\psi({\mathbf{p}}-{\mathbf{k}})}{E-\omega_b({\mathbf{p}}) -
\omega_a({\mathbf{p}}-{\mathbf{k}}) - \omega_\gamma({\mathbf{k}}) } \ \ ,
\end{equation}
however, the self-energy part is divergent:
\begin{equation}
\Sigma_a = \frac{\alpha m_a m_b}{4 \pi^2} 
\int \frac{d^3 {\mathbf{k}}}{\omega_a({\mathbf{p}}-{\mathbf{k}})  
\omega_\gamma({\mathbf{k}})} \frac{1}{E-\omega_b
({\mathbf{p}}) -\omega_a({\mathbf{p}}-{\mathbf{k}}) - \omega_\gamma({\mathbf{k}}) } \ \ .
\end{equation}
The integral is divergent because of the occurrence of the forward part
of the one-loop self-energy correction. We renormalize this
contribution by subtracting the on-shell value:  $\Sigma_r(E) =
\Sigma(E)- \Sigma(\omega_a({\mathbf{p}})+\omega_b({\mathbf{p}}))$.
Renormalization means the redefinition of the parameters, like masses
and charges, of the theory.

We use the on-shell renormalization \cite{Col84}, which boils down to
discarding any change in the values of the physical quantities for the
case of free, constituent particles. This will give the same results in
the same approximation, for observables, as any other approach, but
frees us from lengthy discussions on counterterms and such. However,
there will be small differences in the results, depending whether {\it
on-shell} means on the {\it mass-shell} ($p^2=m^2$), as in the
covariant approach, or on the {\it energy-shell} $(E=\omega$), as
advocated here in the Fock-state truncation.

\begin{figure}
\centerline{\includegraphics[width=8.5cm]{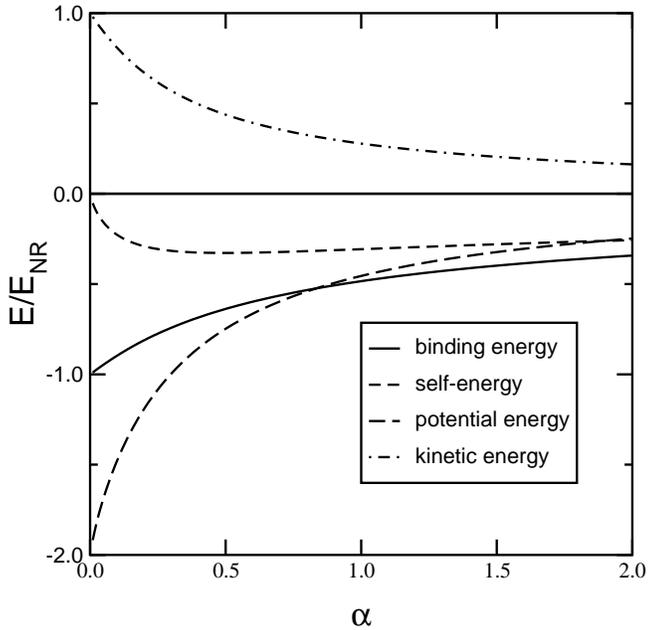}}
 \caption{\label{fig.01}
 The properties of the ground state compared to the
 non-relativistic results, in the case of the masses relevant for the
 hydrogen atom, $m_a = m_p, m_b = m_e$.}
\end{figure}
\begin{figure}
\centerline{\includegraphics[width=8.5cm]{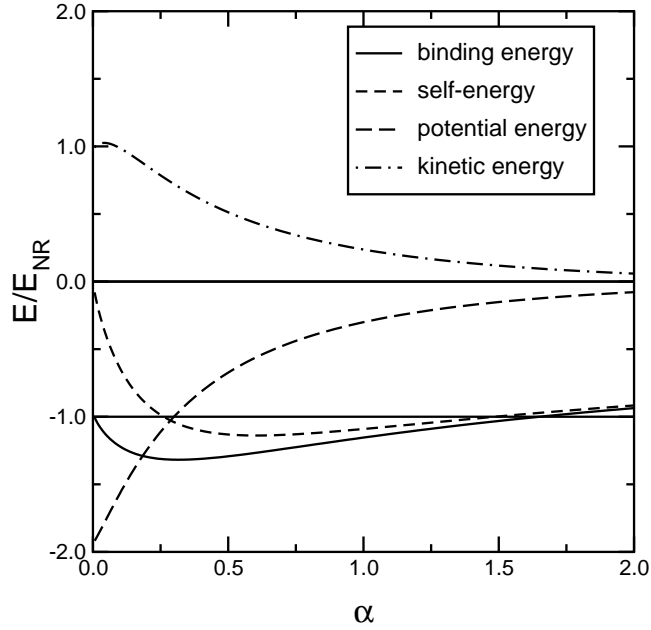}}
 \caption{\label{fig.02}
 The properties of the ground state compared to the
 non-relativistic results, in the case of two equal masses, $m_a=m_b=m$.}
\end{figure}

The angular integrations can be performed analytically.  In order to
calculate the integrals with the least numerical problems we integrate
the self-energy contributions analytically, thereby avoiding the
integrable singularity in the subtraction term
$\Sigma(\omega_a+\omega_b)$ in our numerical calculations. For the sake
of completeness we give the result of a tedious but straightforward
calculation:
\begin{eqnarray}
 \lefteqn{\int_0^\infty
 \frac{k_a {\rm d}k_a }{ \omega_{a}} \ln \left[ \frac{ \omega_{b} -E +
\omega_{a} + |{k_b}+k_a| }{ \omega_{b} -E + \omega_{a} +
|{k_b}-k_a|} \right]_{\rm subtracted}  
 = m_a^2 \ln[m_a]
 \left( \frac{1 }{ {k_b} + \omega_{b}-E} - \frac{1 }{ {k_b} -
 \omega_a} \right)  - }\nonumber \\
 & & { m_a^2 } \ln[{k_b}+ \omega_a] 
  \left( \frac{\omega_a}{ m_a^2} +
 \frac{ \omega_{b} - E }{ (E-\omega_b + \omega_a)(E-\omega_a -
 \omega_b)+m_a^2} \right) + \nonumber \\
 & & \frac{1 }{ 2 } \ln [\omega_a +\omega_b-E] (
 \omega_a + \omega_b-E) 
  \left( 1 + \frac{m_a^2 }{ (\omega_b-E - {k_b} )({k_b}+
 \omega_a)} \right) - \nonumber \\
 & & \frac{1 }{ 2 } \ln \left[ {k_b} + \omega_b -E +
 \frac{ m_a^2 }{ {k_b} + \omega_a} \right]
 ( 2 {k_b} + \omega_a+\omega_b-E)
 \left(1+\frac{m_a^2 }{ ({k_b} + \omega_b-E )({k_b}+ \omega_a)}\right)\, .
\end{eqnarray}
The integrals converge well. With a typical compact integration coordinate
the integrands are smooth bell curves in the middle of the domain, from
which we know that we do not miss any contributions for small or large
values of the momentum.
We use a simple integration procedure and with 500 integration points 
we achieve a sure accuracy of 4 to 5 decimal places, which is in the same 
region as the
systematic error due to the wave-function Ansatz. It is interesting to
observe how different the self-energy contributions are for the
heavy-mass and the equal-mass cases, shown in
Figs.~\ref{fig.01} and \ref{fig.02}, respectively.

Since Hamiltonian renormalization is not a well-established procedure,
although we are convinced of its legitimacy, we repeated the same
calculation with the covariant expression for the self-energy, which means
that only for that part we also include the states with
three particles, one anti-particle, and a photon, which leads to the
known covariant result for the self-energy:
\begin{equation}
 \Sigma_c = \frac{\alpha m^2 }{2 \pi}
 \left(1 - \frac{m^2}{p^2} \right)\ln\left[ \frac{m^2-p^2}{m^2} \right]  \, ,
\end{equation}
\begin{figure}
\centerline{\includegraphics[width=8.5cm]{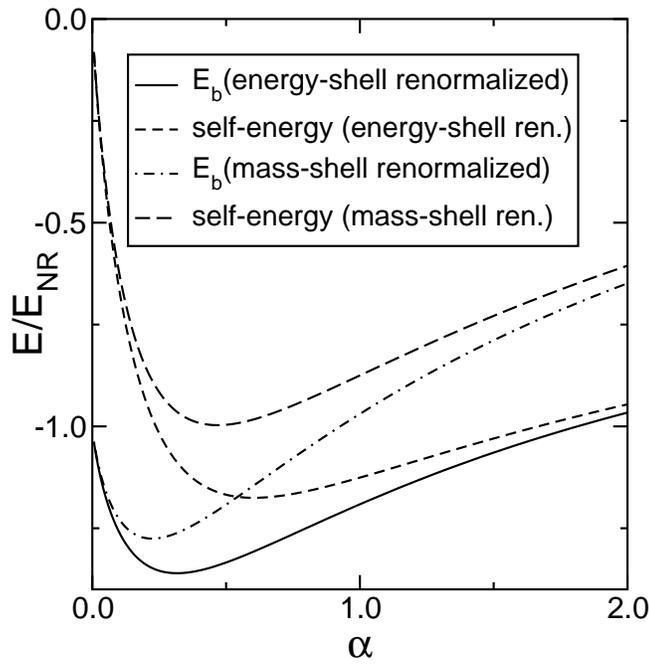}}
 \caption{\label{fig.03}
 The comparison in the equal-mass ground-state binding energies and the
 self-energy contributions to that result 
 between the case where the self-energy is renormalized on energy-shell, 
 which includes only the forward diagram, and the case
 where one-particle mass-shell renormalization, from the corresponding Feynman diagram 
 which also includes the anti-particle contributions, is used. }
\end{figure}
where $\Sigma_c(p^2=m^2) =0$, which corresponds  to the on-mass-shell 
renormalization; for the physical mass there is no mass correction.
For small coupling the result differs only little from the restricted
Fock-state result with $m_a=m_b=m$. For larger values of the coupling
constant the self-energy contribution is slightly suppressed in the 
covariant approach, as can be seen in Fig.~\ref{fig.03}.

The heavy-mass limit yields the correct result, which reduces to the
non-relativistic result in the weak-coupling limit.  The latter problem
is dealt with in a rather cumbersome way in the Bethe-Salpeter equation
and related approaches \cite{Gro93}, as the one-particle exchange
approximation (the ``ladder approximation'') does not work there.

\section{Renormalization}

Since we solve only the lowest non-trivial case of binding with
a massless exchange field, there are many contributions we do not
include. Eventually, to show that the lowest order includes most
of the relevant physics of a bound state, we have to pursue the
problem of convergence. This is part of the future work.
However, since renormalization involves divergent contributions 
already present at the lowest order, also here some care is required.
In the renormalization different orders in the coupling constant are
mixed, while in the Hamiltonian approach some of these terms, which 
should be grouped together, might not appear.

For example, from the Ward identity, it would follow that one
should combine wave-function renormalization of the self-energy
diagrams with the vertex correction, to yield a finite answer.
The former diagram appears in our equation, but the latter do not. 
We choose not to perform the wave-function renormalization, for
one important reason: it leads to a trivial theory, with zero
coupling, in our case of massless exchange particles. In the case
of finite exchange-particle masses the renormalized coupling 
constant scales with the logarithm of the mass. Only if the 
vertex correction is also included, the infinite wave-function 
renormalization is counter-acted by the infinite charge 
renormalization, and only a mild dependence is left. Since 
these divergences are infrared
divergences, it is even more physical\cite{JR76} to group the 
wave-function renormalization with the vertex renormalization,
and have both or none. In our case it is none.

If we compare our result with the work by Ji\cite{Ji94} we find,
as expected, the opposite effect from the self-energy, because
we have chosen the starting Lagrangian in such a way that we
have opposite self-energy contributions; in our Lagrangian identical 
particles repel each other, like they should do in the Coulomb problem.
However, unexpectedly, one can conclude from Ji's work that the
self-energy cures a bit of the illness of the $\phi^3$ theory.
One would expect the binding energy to make a nose dive as the 
coupling constant increases, due to the unboundedness of the Hamiltonian.
However, including the self-energy {\it decreases} the binding
in the case of the scalar exchange as studied by Ji.

For the Bethe-Salpeter equation, with dressed propagators, it 
is found that the self-energy increases the binding slightly, 
however, these results are for a relatively large mass of the
exchange particle, and the wave-function renormalization is used,
which counteracts the weaker binding caused by the mass 
renormalization\cite{AA99}. Moreover, other problems appear
in the covariant Bethe-Salpeter formalism, such as unphysical
solutions\cite{AA99,Nak69}, which have, at least in part, to do with
the relative time-coordinate appearing in the equation. However,
there are claims that these problems result from the unphysical
Lagrangian\cite{RS96}. For the moment we want to conclude that if
renormalization is pursued beyond the simplest on-shell 
subtractions, many questions are still open. The same conclusion
can be drawn from the work of G{\l}azek et al.\cite{GHPSW93} on
light-front Hamiltonian Yukawa theory, which seems, insofar, to confirm
Ji's findings for self-energy effects of real scalar exchange fields.
However, it must be noted that in all these cases the Lagrangian
was different from our Lagrangian, which in practice boiled down to the
opposite sign for the self-energy. This is altogether a different
problem.
When we reversed the sign of the self-energy in
our calculations, as a simple check, we found instabilities for $\alpha >0.5$.
This is a subject for further study.

\section{Conclusion}
Generally, one can say that the approach advocated here is
straightforward compared with the Bethe-Salpeter equation and
subsequent three-dimensional reductions. The normalization is that of
ordinary quantum mechanics, and spurious solutions, associated with
excitations in the relative time do not occur, since the Hamiltonian
framework has only one reference time.  Moreover, issues concerning
anti-particle content of the wave function, present in the covariant
formulation, do not arise here, since the approach is based on the
Fock-state truncation. Another advantage over the standard covariant
approach is the absence of infrared singularities, due to the fact that
the states are always off-energy-shell, and these singularities are
associated with on-energy-shell states with vanishing photon momentum.
\begin{figure}
\centerline{\includegraphics[width=8.5cm]{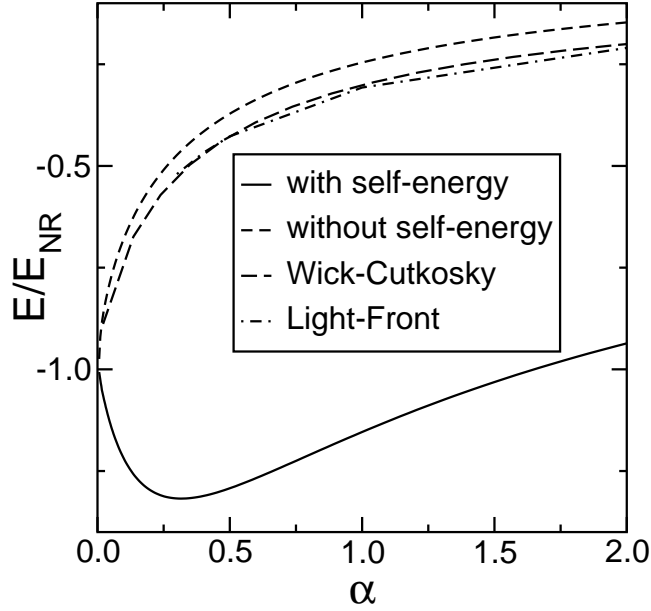}}
 \caption{\label{fig.04}
 Comparison between the result for the full calculation,
 and when the self-energy terms are dropped for the equal-mass case.
We also compare the results with the Wick-Cutkosky model \cite{WC54,Nak69} and
light-front results \cite{MC00}.}
\end{figure}

Apart from the technical advantages of this approach it is important to
notice that the self-energy contributions dominate the ground state in
the strong binding regime.  If we perform the same calculations
including all the relativistic effects except for the self-energy
contribution, we find completely opposite results \cite{MC00}; the
binding energy is lower than the non-relativistic one, as can be seen
in Fig.~\ref{fig.04}. This indicates that the usual description of
strong-coupling binding in terms of exchange diagrams is, to say the
least, incomplete.

\section{Acknowledgments}
We like to thank Prof. John Tjon, Prof. Reinhard Alkofer, and Prof. Franz Gross
for remarks and discussion regarding this work. We also like to thank Jim Friar, Roland 
Rosenfelder, and Axel Weber for their comments.

\end{document}